\documentclass{sig-alternate-10pt}

\usepackage{times,datetime,url,hyperref}
\usepackage{graphicx,multirow,color,calc,ulem,threeparttable,tabularx}
\usepackage{booktabs,enumitem,comment,balance,subcaption}
\usepackage{morefloats}
\usepackage{wrapfig,multirow}
\usepackage[group-separator={,}]{siunitx}
\usepackage{tablefootnote}
\usepackage[font=small]{caption}
\usepackage[draft=true]{minted}

\hypersetup{
  bookmarks=true,
  unicode=true,
  pdftoolbar=true,
  pdfmenubar=true,
  pdffitwindow=true,
  pdfstartview={FitV},
  pdfnewwindow=true,
  colorlinks=false,
  pdfdisplaydoctitle=true,
  pdfborder={0 0 0},
}
\usepackage[all]{hypcap}
\usepackage[absolute]{textpos}

\setlist[itemize]{leftmargin=*,partopsep=0pt,itemsep=0pt,topsep=5pt}
\setlist[enumerate]{leftmargin=*,partopsep=0pt,itemsep=0pt,topsep=5pt}

\textblockorigin{1in}{1in}

\usepackage{tikz}

\usepackage{listings}

\definecolor{dkgreen}{rgb}{0,0.6,0}
\definecolor{gray}{rgb}{0.5,0.5,0.5}
\definecolor{mauve}{rgb}{0.58,0,0.82}

\newcommand{\PhoneLab}{\textsc{PhoneLab}}
\newcommand{\wifi}{Wifi}

\newcommand{\defdroid}{\textsc{DefBot}}
\newcommand{\jouler}{\textsc{Watt}}
\newcommand{\lesson}[1]{\textbf{Lesson:}\space\textit{#1}\vspace*{0.05in}}
\newcommand{\runtime}{\textsc{RTPerm}}
\newcommand{\netsense}{\textsc{NetSense}}

\def\theconference{MobiSys'17}
\def\thetitle{Lessons from Four Years of \textsc{\LARGE PhoneLab} Experimentation}

\def\aftercaptionskip{0mm}

\ifdefined\isdraft
  \usepackage{fancyhdr}
\else
\fi

\begin{document}

\date{}

\title{\thetitle\titlenote{This work is supported in part by NSF grant CNS-1629894.}}

\numberofauthors{1}

\author{
  Jinghao Shi, Edwin Santos and Geoffrey Challen\\
  University at Buffalo\\
  \texttt{\{jinghaos, edwinsan, challen\}@buffalo.edu}
}

\CopyrightYear{2017}
\conferenceinfo{MobiSys'17}{June 19-23, 2017, Niagara Falls, USA}
\crdata{}

\hypersetup{
  pdfinfo={
    Title={\thetitle},
  },
}

\maketitle

\begin{abstract}

Over the last four years we have operated a public smartphone platform
testbed called \PhoneLab{}.
\PhoneLab{} consists of up to several-hundred participants who run an
experimental platform image on their primary smartphone.
The experimental platform consists of both instrumentation and experimental
changes to platform components, including core Android services and Linux.

This paper describes the design of the testbed, the process of conducting
\PhoneLab{} experiments, and some of the research the testbed has supported.
We also offer many lessons learned along the way, almost all of which have
been learned the hard way---through trial and a lot of error.
We expect our experiences will help those contemplating operating large
user-facing testbeds, anyone conducting experiments on smartphones, and many
mobile systems researchers.

\end{abstract}
 \section{Introduction}
\label{sec:intro}

Smartphones are the most successful mobile devices in computing history.
Almost 2~billion of these powerful devices are already deployed worldwide.
For many---particularly in developing countries---a smartphone is their only
computer and only internet connection.
Smartphones provide a rich set of opportunities and challenges for mobile
systems researchers as we adapt to a mobile-first world.
\sloppy{
The global network of distributed smartphones represents the ultimate proving
ground for experimental approaches to crowdsourcing, wireless
communication, location tracking, energy management, context awareness, security
and privacy, user interfaces, and other topics of interest to the mobile systems
research community.
}
Smartphones are the ultimate destination for many of our new ideas.

A key challenge when transitioning ideas from local laboratories to the
global stage is determining whether they work for large numbers of users.
Initial experiments tend to be done by researchers themselves.
They are inherently small scale, and use participants that are not
representative of typical smartphone users.
From an initial small-scale prototype with a handful of sophisticated users,
it is a big jump to transition to billions of unsophisticated users.
An intermediate step would help determine whether ideas that initially seem
successful are truly ready for widespread deployment.

If a new idea can be evaluated through a smartphone app, experimenters can
use smartphone software marketplaces to perform medium-scale studies.
Researchers first integrate their new idea into an app---ideally a useful
app.
Then they deploy it on an app marketplace like the Google Play Store.
Recruiting several hundred users to install and use it can be done through a
mixture of advertising and participation incentives.
Or they may simply count on a combination of human curiosity and the enormous
numbers of users they can reach through app marketplaces.
Given a large enough audience, even a tiny amount of interest can compound to
enough participants to complete a study.
Previous experiments have successfully utilized this approach to recruit
several hundred participants for a variety of
projects~\cite{carat-sensys13,ubicomp2014-pocketparker}.
And, when you mix a useful app that meets a common user need with a bit of
timely publicity, the results can be explosive.
After receiving some good press, the Carat energy management tool was
eventually installed by almost 1~million users~\cite{carat-sensys13}.

But not every new idea can be deployed as an app.
Smartphones also run a complex million-line codebase referred to as the
smartphone platform, which provides the interface used by apps.
On Android the smartphone platform consists of three main components.
The Android SDK is built along with the platform into
\texttt{android.jar} and utilized by all apps.
This code runs in the app's process context and implements many core Android
features as well as communication with Android services.
Core Android services---such as the \texttt{LocationManager}---run as
separate processes and communicate with apps to provide information.
The Linux kernel performs typical operating system functions such as
scheduling and other forms of resource management.
Apps rely on---but cannot modify---these platform components.

Because the platform provides functionality needed by many apps, it is also a
natural location for mobile systems research and experimentation.
Among other responsibilities, the platform estimates
location~\cite{roy2014smartphone,liu2013guoguo,nandakumar2012centaur,peng2007beepbeep}, manages
energy~\cite{mobicase2015-jouler,carat-sensys13,xu2013optimizing,ding2013characterizing}, chooses
between networks~\cite{infocom2016-scans,deng2014wifi}, and attempts to secure the
device~\cite{mobicase2014-pocketlocker,defdroid,lockscreen,mirzamohammadi2016viola}.
All these are areas of ongoing mobile systems research.
But that want to evaluate improvements to these core features cannot rely on
the app store to perform medium-scale experimentation.
They require another way to perform their experiments.

\begin{figure*}[t]
	\centering
	\includegraphics[width=\textwidth]{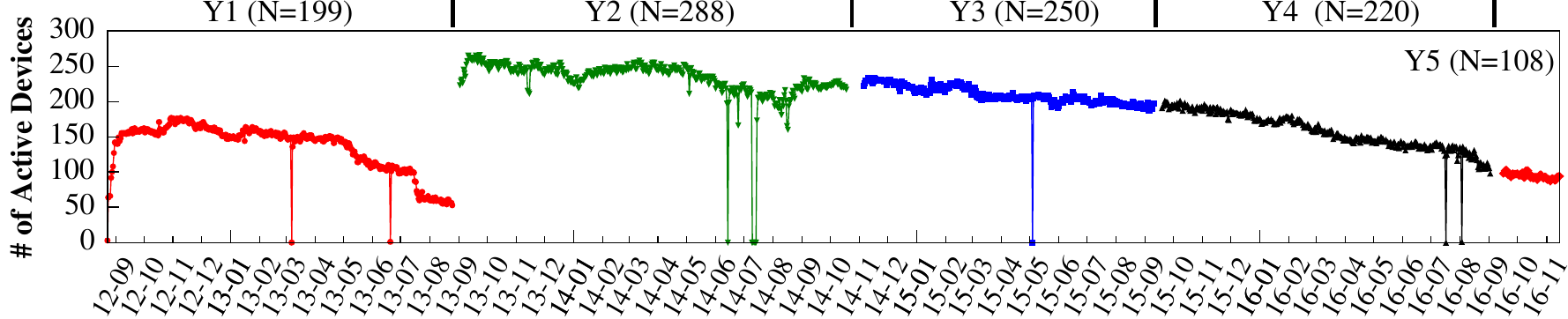}
	\caption{\textbf{Daily Active Devices on \PhoneLab{}}.
    Transition periods between operation years are not shown.
    There were several sudden drops that were caused by backend server
  maintenance.
  The participant counts shown above the figure are when the operation year
started.}
	\label{fig:active}
  \vspace*{\aftercaptionskip}
\end{figure*}

For the past four years we have operated a public smartphone platform testbed
called \PhoneLab{}.
\PhoneLab{} allows researchers to deploy platform modifications to between
100~and~300 participants.
Participants use their \PhoneLab{} smartphone as their primary smartphone,
and so provide representative usage patterns.
Participants are incentivized by a free device, a low-cost service plan, and
on-campus technical support.
They agree to use experimental software and understand that data will be
collected from their device.

Any researcher can distribute experiments on \PhoneLab{} free of charge.
We have attempted to make the process as straightforward as possible by
providing detailed instructions.
But \PhoneLab{} experimentation does require modifying the Android Open
Source Project (AOSP) sources---a daunting task for the uninitiated.
\PhoneLab{} experiments typically fall into one of two categories.
New instrumentation can be added to the platform to generate data about how
the smartphone is used.
This is a common first step in understanding a problem and beginning to
formulate a solution.
Instrumentation usually outlives the experiment that introduced it, and our
\PhoneLab{} codebase includes an increasing amount of useful logging.
When a new feature is ready to be tested, it can also be deployed on
\PhoneLab{}---usually along with additional instrumentation required to
evaluate it.

Operating \PhoneLab{} has been both fun and challenging.
The goal of this paper is to describe the testbed, advertise available data
sets, and share what we have learned.
Some of our experiences echo those of others that have built and maintained
computer systems.
But others are specific to the challenge of running a \textit{user-facing}
testbed.
\PhoneLab{} is more than a set of machines locked in a room~\cite{emulab},
tucked into faculty offices~\cite{motelab-ipsn05}, or distributed around the
world~\cite{chun2003planetlab}.
It is a set of people that rely on their smartphone in the same ways as the
rest of us.
This creates unique design and operational challenges, particularly given
that we allow experimenters to modify low-level platform code.
If an experiment crashes a typical testbed, someone may miss a paper
deadline.
If an experiment crashes \PhoneLab{}, someone may not be able to call 911.
We have also encountered unique challenges associated with incentivizing and
interacting with participants, providing safe access to data collected from
human participants, and dealing with cellular service providers.

\begin{comment}
The rest of our paper is structured as follows.
We begin in Section~\ref{sec:history} by describing the history of
\PhoneLab{} from conception to its current realization.
Section~\ref{sec:op} describes how \PhoneLab{} experiments are performed.
To highlight how \PhoneLab{} benefits mobile systems experimentation, we
describe several sample experiments in Section~\ref{sec:exp}.
Section~\ref{sec:dataset} provides an overview of the dataset we have
collected and make available to interested scientists.
We describe our experiences and lessons learned from running \PhoneLab{} in
Section~\ref{sec:lesson}, and continue the discussion in
Section~\ref{sec:discussion} by considering the future of smartphone
testbeds.
We describe related efforts in Section~\ref{sec:related} before concluding in
Section~\ref{sec:conclusion}.
\end{comment}
 \section{The PhoneLab Testbed}
\label{sec:history}

In this section, we provide an overview of \PhoneLab{}'s history
(\S\ref{subsec:history}), hardware and software components
(\S\ref{subsec:hardware}), privacy and safety measures
(\S\ref{subsec:privacy}) and the transition from app to platform experimentation
(\S\ref{subsec:transition}).

\subsection{History}
\label{subsec:history}

Figure~\ref{fig:active} shows the daily \textit{active} devices during the 4
years of \PhoneLab{} operation.
The data is obtained from the smartphone heartbeat information received by
our backend server.
The daily active count is a rough estimate but strictly smaller than the
total number of participants.
Several sudden drops of active count are caused by maintenance or problems
with our backend data collection server.

\textbf{Year 1 (9/2012--9/2013):}
We recruited the first group of 199 participants in September 2012 by
offering a free smartphone and a free year of service.
Participants were supposed to transition to a paid plan in their second year,
but received one year of free service as a recruitment incentive.
The first year served as beta-testing of our platform software and data
collection framework.

\textbf{Year 2 (9/2013--11/2014):}
We recruited a second group of 288 participants in September 2013.
Although we had hoped that they would continue with the project, most of the
Y2 participants were new and most of the Y1 participants left the project.
This was caused by poor incentive design, a topic we return to later in
Section~\ref{subsec:incentives}.
\PhoneLab{} opened to the public for app-based experiments in October 2013.

\textbf{Year 3 (11/2014--9/2015):}
The third year of \PhoneLab{} brought changes to both the incentive model and
experimental capabilities.
We eliminated the free service plans which were not functioning as an
effective incentive for recruiting high quality participants.
Instead, we began billing participants at a discounted rate for their
cellular service.
We began running platform experiments privately early in 2014, and made this
capability available to the public in Feb. 2015.

\textbf{Year 4+ (9/2015--Present):}
Not providing free service lowered the cost of the testbed dramatically and
allowed us to continue to run it for another fourth year on three years of
initial funding.
Given uncertainty about future funding, we stopped recruiting participants in
Y4.
Attrition began to decrease the size of the testbed.

\begin{comment}
In Y5, which began September 2016, we received a small amount of additional
funding for research staff but no additional funding for participant service
plans.
Instead, the threat of closing down the project entirely allowed us to
renegotiate a slightly better rate with Sprint.
We also raised the participant monthly cost from \$45 to \$50.
These two changes combined to generate enough extra revenue to purchase new
devices and continue paying our undergraduate administrator who was no longer
funded out of the grant.
\end{comment}

\subsection{Hardware and Software}
\label{subsec:hardware}

We used Google Nexus devices exclusively through the \PhoneLab{} project.
We began with the Nexus~S in Y1, and then upgraded to the Galaxy Nexus (Y2),
Nexus 5 (Y3--Y4), and Nexus 6 (Y5).
We chose the Google Nexus series devices because of its developer support and
driver availability.

\begin{comment}
For each device generation, the first task was to build a fully-functional
platform image from the latest set of AOSP sources.
This is supposed to be possible for the Nexus series of smartphones.
But despite official binary driver availability, this was not a trivial task.
Each year brought new problems with various devices components, including the
camera, cellular radio, and GPS chipset.
\end{comment}

Once we had a fully-functional Android ROM as a baseline, we made three
important modifications to support experimentation.
First, we developed a testbed management app called the
\PhoneLab{} Conductor.
The Conductor is built into the platform and is configured to start on boot.
It serves to link each smartphone with our backend server.
It sends heartbeats, collects and uploads data, and fetches experiment and
platform updates.

Second, we modified the Android \texttt{logcat} system to facilitate
\PhoneLab{} data collection.
The changes include enlarging the in-memory log buffer, increasing the
maximum characters that can be logged per line, and improving the timestamp
accuracy from millisecond to microseconds.
These are very important as we rely heavily on \texttt{logcat} and the
metadata it attaches when processing logs.
We discuss the detailed data collection mechanism in
Section~\ref{subsec:collect}.

Finally, we added instrumentation to Android to collect information useful to
many experiments: location updates, Wifi scan results, battery status
changes, etc.
All such instrumentation is done passively by piggybacking on places where
Android already collects this information.
We have been careful to avoid adding any timers or other forms of overhead to
our instrumented platform.
A full and detailed list of instrumentation can be found on our
website\footnote{Link omitted for double-blind review.}.
The entire \PhoneLab{} data set is always available to interested
researchers, subject to human subjects review.
We describe the dataset and data release process in
Section~\ref{sec:dataset}.

\subsection{Participants}
\label{subsec:incentive}

Almost all \PhoneLab{} participants are faculty, staff and students at our
university.
When initially providing a free device and service as an incentive in Y1 and
Y2, we required a university affiliation in case we needed to recover the
device.
Once we began billing participants in Y3, that made it possible to recruit
from outside the university.
If participants stop paying their bills, we simply cancel their service, and
so no longer require any additional enforcement mechanisms.

However, we have found it convenient to continue recruiting university
participants.
It has the effect of making the testbed denser, so most participants spend
their working hours on our fairly compact campus.
It also allows us to advertise on-campus technical support as an attractive
non-monetary incentive.
Participants seem to appreciate that they can drop by our lab during office
hours for help, rather than having to make a special trip to a cellular
store.
Despite being all university affiliates, our participants are a diverse
group---both in terms of age, gender, and occupation.

\subsection{Privacy Concerns}
\label{subsec:privacy}

Smartphones are personal devices, and can potentially reveal a great deal of
sensitive information about their user.
When operating our testbed we must take steps to protect the privacy and
safely of \PhoneLab{} participants.
This is particularly true given that we are now distributing platform
updates.
Android places limits on what apps can do.
But the platform has no such restrictions.

First, all \PhoneLab{} personnel---including faculty, developers, and
administrators---are required to complete human subjects training.
At our university this includes a Good Research Practices (GRP) course and
the Social and Behavioral Research Investigators training provided by the
CITI program~\cite{citi}.
These programs ensure that the \PhoneLab{} team understands the importance of
protecting our participants and best practices for handling \PhoneLab{} data.

Second, we apply standard security techniques to protect data during
collection and storage.
\PhoneLab{} logs are transmitted to the backend server over HTTPS.
The server is located in a secured server room in our department.
Access to the server is granted to only a handful of \PhoneLab{} developers
who handle data collection.

Finally, all researchers requesting \PhoneLab{} data are required to have
their request reviewed for human subjects safety.
At US universities, this is done by the Institutional Review Board (IRB).
Non-academic institutions and universities outside the US typically have an
equivalent body.
This is standard practice in any projects that involves human subjects.
We are pleased to see IRB review increasing expected by the mobile systems
community.
We discuss our experiences with the IRB in more detail in
Section~\ref{subsec:irb}.

\subsection{Experiment Model}
\label{subsec:transition}

During Y1 and Y2 of \PhoneLab{} we focused on supporting app-level
experiments.
Researchers would publish their app on the Google Play Store, and we would
notify \PhoneLab{} participants and ask them to install and use it.
However, it was challenging to determine whether users were actually
participating in these experiments.
It was possible to monitor whether they installed an experiment and spent at
least some time with it in the foreground.
But whether they were using it realistically was difficult to determine.
It also became apparent to us that \PhoneLab{} would never provide as many
participants as app-level experiments could reach through app marketplaces.

As a result, we began to shift to platform experimentation.
This is not only a unique capability that \PhoneLab{} was able to provide, but
also made it easier to track participation.
For many platform experiments, users participate silently without having to
install or use any new software.
We do notify users of new platform experiments and provide them with ways to
limit data collection on a per-experiment basis.
Starting from Y3, \PhoneLab{} provided only platform experimentation
capabilities, and we remain the only public testbed to do so.
 \section{Testbed Operation}
\label{sec:op}

In this section, we describe the experiment workflow and operation of
\PhoneLab{}.
In particular, we focus on the trade-offs we made trying to protect
\PhoneLab{} participants from buggy experiments while making \PhoneLab{}
operation as transparent to participants as possible.

\subsection{Experiment Workflow}

\PhoneLab{} experimentation proceeds in three phases: creation, development and
deployment.

\subsubsection{Creation}

Researchers who are interested in \PhoneLab{} first contact us and describe
the scope and purpose of the experiment.
We review this information based on two criteria: suitability and
intrusiveness.
Since Y3 \PhoneLab{} has focused on platform-level experimentation.
Experiments that can be deployed as apps are encouraged to utilize other
approaches such as the Google Play store.

We also evaluate how disruptive the experiment will be to the \PhoneLab{}
participants.
Experiments that require specific user behaviors or incur large performance
or battery penalties will not be deployed.
Over the years, we have been impressed, however, with the kinds of
experiments that can run without disturbing users.
Several times, logging and instrumentation that we were initially worried
would caused too high overhead ended up being unnoticeable.
As an example, we have instrumented the system call interface to collect file
system activity traces.
This generates a great deal of output, but has been run successfully without
complaint from participants.

Note that IRB approval is not required at this stage.
But experimenters are made aware that the IRB letter is necessary for the
data collection after deployment.
This lowers the barrier for experimenting on \PhoneLab{}, allowing the
experimenters to work on the development and IRB application in parallel.
The experiment creation stage usually takes just a few days.

\subsubsection{Development}

\sloppy{
After approval, dedicated experiment branch is forked from our AOSP code base
where the researchers can stage their modifications.
Once experimental changes are ready, a \PhoneLab{} team member checks out the
branch and bench tests it on a single device.
Assuming that is successful and no glaring problems are immediately
apparent, the modified platform is pushed to a small number of \PhoneLab{}
developers.
Anomalies such as app crashes, performance degradation, and unexpected
battery drain are reported at this stage.
If problems arise, experimenters are notified and asked to fix them.
This step avoids causing problems for larger numbers of participants that are
not developers, and has dramatically reducing the burden of our technical
support.
The development-testing iteration repeats until all problems detected by
\PhoneLab{} developers are fixed.
}

\subsubsection{Deployment}

After testing is complete, changes are pushed to testbed participants.
Depending on the amount of time requested by the experimenters, this stage
can last for weeks, months or years.
After the requested period ends, we remove the experimental changes from
participant devices.
However, if experimenters have added generally-useful instrumentation, we
encourage them to merge it into our master development branch.
This allows logging for that instrumentation to continue and makes data sets
available later to other researchers.

Any researcher can request \PhoneLab{} data sets with IRB approval, and so
this process is decoupled from the experimentation workflow.
However, most experiments end by researchers requesting a data set containing
tags generated by their experiment and other useful information.
To assist experimenters during iteration, we can provide small data sets for
the purposes of validating their instrumentation without receiving IRB
approval.
IRB approval is only required once experimenters plan to publish their
results.

\begin{comment}
\subsection{Branching Philosophy}

\PhoneLab{} developers and experimenters collaborate on the same platform
code but with different goals.
Developers aim to provide a usable smartphone platform and maintain common
instrumentation.
Experimenters often want to augment the Android platform in various ways.
Due to the unstable nature of experimental changes and performance and energy
overhead they can produce, we need the ability to remove experiments after
they are deployed.
As a result, the canonical topic-branch and patch model of AOSP development
does not meet our needs.
It renders difficult if not impossible to remove a changeset after it is
merged into the main deployment branch.

We work around this AOSP limitation with a new development model.
All experiment branches are forked from a common ancestor, which we refer as
the ``big-bang''.
Every time a platform release is scheduled, a new release branch is forked
from ``big-bang''.
Active experiment branches are then merged into this release branch in the
order they were created.
Later experiments that generate conflicts with earlier experiments are
skipped.
The resulting release branch is then built and pushed to participants.
This allows experiments to remain isolated in their respective branch.
It also allows us to remove a deployed experiment from the participant
devices by skipping the branch during the next release.
\end{comment}

\subsection{OTA Updates}

As a smartphone platform testbed, the most important capability is to push
platform changes to participant devices.
We leverage Android's existing over-the-air (OTA) update mechanism for this
purpose.
It provides a way to initiate an update to a new platform by applying an
update file stored locally.
Our job is to generate the update file and reliably transmit the update file to
the device.
Every time a new platform image is built, incremental OTA update packages are
generated against previous platform versions.
The \PhoneLab{} Conductor app periodically checks for OTA updates from the
backend server, downloads them, and prompts participants to install the
update once the download completes.
To ensure that updates are eventually applied, the Conductor will
automatically apply a pending OTA after midnight once the phone is charging
and is not interactively used.

\subsection{Data Collection}
\label{subsec:collect}

Android has a built-in \texttt{logcat} system that allows apps and various
parts of the framework to log debug messages or events.
Many useful contextual information are already being logged by the framework,
such as screen status, \wifi{} connection status, battery level, etc.
All messages are stored in an in-memory ring buffer.
To harvest this information, we harness \texttt{logcat} as a sink for our
device-end data collection.
Experimenters are instructed to log their data using the common \texttt{logcat}
interface.
\begin{comment}
For security reasons, \texttt{logcat} instances are normally limited to
collecting logs from the app that spawned them.
Otherwise \texttt{logcat} would leak potentially sensitive information
between apps.
But because the Conductor is signed to match the platform image, the
\texttt{logcat} instances it starts can collect logs from all apps.
\end{comment}

We also developed utilities to pipe the Linux kernel logs and the kernel
event tracing logs~\cite{tracing} to the \texttt{logcat} buffer.
This allows us to hook into the existing Linux logging framework.
Listing~\ref{lst:log} shows examples on how to instrument various parts of the
platform.

\begin{lstlisting}[
  language=Java,float,floatplacement=t!,
  caption={\textbf{Various Utilities for Instrumetation.} All log messages are
  piped to the \texttt{logcat} buffer.},
  label={lst:log},
  belowskip=-3mm
]
// Java: use the Log class.
Log.d("MyTag", "Some useful information.");

// C/C++ (user space): use Android ALOG macros.
ALOG(LOG_INFO, "MyTag", "Another useful log.");

// Kernel: use printk and tracing framework
printk(KERN_INFO "Useful kernel logs.");
trace_my_tracepoint(...);
\end{lstlisting}

The \PhoneLab{} Conductor app constantly consume logs from the buffer and
dumps them into log files.
These files are then uploaded to our backend server.
Similar to OTA updates, the data upload also happens in a opportunistic way
when the device is charging.

Once the backend server receives the log files, it performs a series of
processing steps to clean them up and add additional information.
The resulting log files include a hashed device identifier and the upload
time in addition to the original data generated on the device by
\texttt{logcat}.
Logs are sorted into a set of flat files stored on a large RAID array.
We currently store one file per day per device containing all generated log
messages.
To process a data request, we filter the full set of log files by time and by
tag to produce the data that is returned to the experimenter.
 \section{Example Experiments}
\label{sec:exp}

\sloppy{
Over the last two years as a smartphone platform testbed, \PhoneLab{} has
accommodated 17 experiments: 8 from \PhoneLab{} developers and
9 from external researchers.
}
In this section, we showcase four experiments\footnote{Information that may
reveal author identities, including project names and citations, are
concealed for double-blind review.} that were deployed and evaluated on
\PhoneLab{}.
We demonstrate how \PhoneLab{} enables researchers to examine real smartphone
behavior at scale, and deploy platform changes that cannot be distributed any
other way.

\subsection{Defensive Mobile OS}

\defdroid{}~\cite{defdroid} researchers studied disruptive app behaviors:
waking up devices too often, overuse of GPS, and frequent notifications.
Disruptive behaviors were detected by tracking the apps's API calls to
certain Android platform services, such as \texttt{AlarmManager} and
\texttt{LocationManager}.
Defensive actions were enforced by modifying Android to hijack and alter API
calls corresponding to disruptive behavior.
It is clear that neither the detection nor the defensive actions can be
achieved without modifying the Android platform.

\begin{comment}
\defdroid{} experimenters worked with the \PhoneLab{} to ensure that
defensive actions did not break apps.
This is also a key evaluation criteria for the \defdroid{} approach.
First, we closely monitored complains about app misbehaviors during the
experiment, and forwarded them to \defdroid{} experimenters for examination.
Second, we ask \defdroid{} experimenters to provide a per-app toggle in the
Android settings to set \defdroid{} to permissive mode, in which disruptive
behaviors will be logged but defensive actions will be taken.
This allowed participants to disable \defdroid{} for certain apps if they
thought it was negatively affecting the app's behavior.
We communicated instructions on how to do this to \PhoneLab{} participants
when the \defdroid{} experiment was deployed.
\end{comment}

\defdroid{} was deployed on \PhoneLab{} for over a month from 09/21/2015 to
11/03/2015.
The \defdroid{} experimenters acknowledged that real deployment on
\PhoneLab{} was helpful to verify that the defending actions did not break
the functionality of large numbers of apps used by real users.
They also identified the deployment as an opportunity to collaboratively tune
defensive policies using feedback from \PhoneLab{} participants and
developers.

\subsection{Lock Screen Analysis}

This experiment studied the protection mechanism that users enabled
on their smartphone lock screen.
The Android framework was instrumented to transparently log
the type of each unlock events.
Detailed unlocking behaviors, such as time spent on unlocking the screen,
number of attempts before a successful unlock, and correlation between PIN
length and unlock time, were also logged to study the user interaction with
the unlocking mechanism.
This information can not be accessed by apps for security reasons, but can be
easily logged by modifying the Android platform.

This experiment was deployed on \PhoneLab{} for 8 months from 10/22/2015 to
06/03/2016.
Based on the data collected, the experimenters validated
the correlation between the user's \textit{securement} sub-scale of Security
Behavior Intention Scale (SeBIS)~\cite{egelman2015scaling} score and the
smartphone lock screen usage~\cite{lockscreen}.
Experimenters also identified areas where the lock screen mechanism can be
improved to increase usability while still maintaining
security~\cite{harbach2016anatomy}.

\subsection{Smartphone Energy Manager}

The \jouler{}~\cite{maiti2015jouler} experiment aimed to separate the energy
management mechanism from policy on smartphones.
A new Android framework service was developed to add APIs that provide the
measurements of per-app energy consumption.
It also added new mechanisms to tune app energy consumption via methods such
as CPU frequency throttling.
These knobs enable user level energy managers to enforce different management
policies, rather than relying on policies baked in to the platform itself.

The experimenters also provided several example policies to demonstrate how
the \jouler{} framework can be used to write flexible user space energy
management policies.
Participants were notified the availability of such policies and were
encouraged to choose the one they prefer.
The \jouler{} experiment was deployed on \PhoneLab{} for over a week from
03/07/2016 to 03/16/2016.
An exit survey was also distributed at the end of the study to collect
subjective feedback from participants.

\subsection{Runtime Permission Model}

The \runtime{}~\cite{baokar2016contextually} experiment instrumented the
Android framework to record resource requests made by application that are
protected by installation time permission manifests, such as location,
storage, camera, sending SMS, etc.
Upon detection of such events, a dialog appears asking the user whether the
request would have been declined had they been given a choice.
The experiment itself did not block resource access but only recorded the
participant's choice.
As a result, it does not affect app functionality.

Based on data collected on \PhoneLab{} participants, the experimenter
proposed a runtime permission model where such resource permission shall be
granted at runtime when they are requested instead of statically declared in
the application's manifest at installation time.
In fact, this is similar to the permission model that was adopted in later
Android versions beginning with 6.0.
The \runtime{} experiment was deployed on \PhoneLab{} for about 4 months from
11/24/2015 to 03/16/2016.
 \section{\PhoneLab{} Dataset}
\label{sec:dataset}

Next, we first briefly describe the available dataset generated by
\PhoneLab{}, and the process of data release.
Collected data falls into three categories: existing Android logs,
instrumentation added by \PhoneLab{} developers, and temporary
instrumentation added by experimenters.

\subsection{Android \& Linux Logs}

Android's \texttt{logcat} is intended to be used for debugging.
During development, app developers can use the \texttt{Log}~\cite{log}
utility to print certain debug information.
Upon app crashes, the relevant logs together with stack traces will be upload
as part of the crash report.
According to the Android development guide, developers are instructed to
remove debugging logs in production code.
Yet we still found plenty of app-specific log messages in our dataset.

Additionally, logs are naturally generated by many core Android services as
well: including the Dalvik and Art Java runtimes, the \texttt{SurfaceFlinger}
display rendering engine, the core Android UI framework, and other core
components.
Finally, we also pipe the Linux kernel logs generated by \texttt{printk} to
the \texttt{logcat} buffer.
Taken together, this subset of our data represents a unique opportunity to
study Android behavior in the wild.

\subsection{\PhoneLab{} Instrumentations}

As part of \PhoneLab{} platform development, we instrumented the Android
framework to log various useful information: battery status/level changes,
location updates, \wifi{} scan results, etc.
Most of the information can be also obtained from the apps using the Android
API.
However, we can usually obtain more information inside the framework.

For instance, the Android \texttt{TelephonyProvider} service internally
records the detailed cellular tower and signal strength information, such as
LTE RSRP/RSRQ/CQI.
Yet most of this information is not provided to apps.
Platform instrumentation allows us to add logging at the place where all
available information is still preserved.

\subsection{Experiment Modifications}

Besides augmenting the Android framework with new features, most experiments
also added instrumentation to either motivate or validate their experiment.
We believe the data can also be potentially useful for other researchers with
similar interests.
Examples of such information include SQLite queries, file system access
patterns, lock screen behaviors and app energy consumption details.

\subsection{The Dataset}

Over the 4 years of \PhoneLab{} operation, we collected 148 billion log
lines totaling 4.6~TB of compressed log files.
Each log line contains 6 fields: device ID, timestamp, Linux task ID, log
level, tag, and message body.
The device ID is a hashed string of the device's MEID~\cite{meid}, and is
guaranteed to be unique to each device and consistent across the dataset.
The timestamp field contains the Unix timestamp (with microsecond accuracy)
when the log line was generated.
The rest of the fields are the same with the \texttt{threadtime} format for
Android's \texttt{logcat} system.
The tag is a string assigned by the developer to identify the purpose or of
the log line.
Android does not pose any constraints on the format of the message body.
To simplify post-processing, we require all logs added by \PhoneLab{}
developers or experimenters to use the JSON format for log messages.

\begin{table}[t]
	\centering
	\begin{tabularx}{\columnwidth}{Xrr}
		Category & Tag Count & Line Count \\\midrule
		\PhoneLab{} & 55 & \num{8674766791} \\
		Experiments & 58 & \num{2471169521} \\
		Other & \num{12691} & \num{137714283583} \\\midrule
		Total & \num{12804} & \num{148860219895}\\
	\end{tabularx}
	\caption{\textbf{Overview of \PhoneLab{} Dataset.}}
	\label{tab:dataset}
  \vspace*{\aftercaptionskip}
\end{table}

Table~\ref{tab:dataset} shows a breakdown of the dataset in each of the three
categories.
Tags added by \PhoneLab{} developers or experimenters are recognized by an
enforced naming convention, which contains the experiment code name and
author institution identifier.
All tags not recognized by the convention are categorized as ``Other''.
\PhoneLab{} instrumentation is included in every platform update, while
experiment logs are only deployed temporarily.
So while certain experiments can generate large volume of logs---file system
accesses, SQLite queries---the total number of experiment logs is less than
the \PhoneLab{} instrumentation.

\begin{table}[t!]
  \centering
  \begin{tabularx}{\columnwidth}{Xrl}
    Tag & Line \# & Description\\\midrule
    \texttt{Kernel-Trace}               & 52.7B & Linux event tracing logs.\\
    \texttt{SurfaceFlinger}             &  6.1B & Android rendering.  \\
    \texttt{dalvikvm}                   &  5.4B & DalvikVM.\\
    \texttt{MP-Decision}                &  1.8B & CPU hotplug.  \\
    \texttt{art}                        &   0.3B & ART VM (after Lollipop).  \\
  \end{tabularx}
  \caption{\textbf{Example Tags Generated by Android Framework and Linux Kernel in \PhoneLab{} dataset.}}
  \label{tab:top}
  \vspace*{\aftercaptionskip}
\end{table}

Table~\ref{tab:top} shows several example tags from the ``Other'' category.
The \texttt{Kernel-Tracing} tag includes the log messages from the Linux
kernel event tracing~\cite{tracing} framework.
Useful informations such as CPU/GPU frequency changes, context switch and
scheduling, CPU temperature changes, are logged under this tag.
The \texttt{SurfaceFlinger} tag is generated by the Android
\texttt{SurfaceFlinger} framework, which handles the actual UI and graphics
drawing.
Information such as frame rate and inter-frame intervals can be obtained to
infer UI smoothness and user experience.
The \texttt{dalvikvm} and the \texttt{art} tags are generated from the Dalvik
Java VM (prior to Android 5.0 Lollipop) and the new ART VM.
Existing \texttt{dalvikm} logging exposed information about garbage collector
activity.
Finally, the \texttt{MP-Decision} tag is generated by the proprietary CPU
hotplug feature found on Qualcomm chips.
It records the status of cores as they are brought online and offline as
needed.

\subsection{Data Release}
\label{subsec:release}

We are working on an online catalog of the \PhoneLab{} dataset.
Once complete, it will allow researchers to browse the available data and
decide the list of tags and time range of interest.
An IRB approval is required to submit a data request.
Since most of the data does not contain any personal identification
information, we expect IRB exemption in most cases.
After reviewing the IRB letter, we will collect and provide the requested
subset of the data.
 \section{Lessons Learned}
\label{sec:lesson}

We have learned a great deal from building and operating \PhoneLab{}.
Below we attempt to distill some of the lessons we have learned along the
way.
Some of the lessons will not surprise those that have built and maintained
large computer systems, particularly user-facing ones.
We hope that these lessons are valuable to anyone in the mobile systems
community contemplating experimenting with human subjects.

\subsection{Managing Participants}

\PhoneLab{} is nothing without the human participants that choose to
participate in the project.
And so our first set of lessons concerns how to recruit and manage human
participants.
This is a challenge not faced by equipment-only testbeds such as EmuLab,
PlanetLab and MoteLab.
A computer does not decide that it wants to quit the testbed and join some
other project.
And a computer does not get angry because it came by for help five minutes
after office hours ended and the \PhoneLab{} administrator had already left.
Given a fixed amount of resources---money or staff time---the goal is to
jointly maximize the size and quality of the participant pool.
With this goal in mind, we offer the following recommendations.

\subsubsection{Get the Incentives Right}
\label{subsec:incentives}

\lesson{When you give someone something for free, they are free to treat it
like it has no value.}

During the first year of \PhoneLab{} participants were offered a free
smartphone and free cellular service---including unlimited text, talk, and
data.
The original plan was that we would incentivize participants with a free year
and then move them to a revenue-neutral paid plan for subsequent years.
As a result, a large amount of the \$1M grant from the NSF was dedicated to
paying cellular service fees to Sprint to cover project participants.

As you would expect, it was \textit{very} easy to give out 200~free phones
with free service.
Everyone loves free stuff.
And it was easy on our \PhoneLab{} administrators and staff.
There was no billing to do, and because participants were not paying anything
their customer service expectations were low.

We had been careful to instruct participants to use the \PhoneLab{}
smartphone as their primary device.
But as we began examining the data, we realized that many of them were not.
When the year ended and we began to retrieve devices, some were still in
their original shrink wrap.
Other participants were clearly just using the device for data tethering, or
to experiment with Nexus smartphone and test out Sprint's service.
Overall the free service incentive made it easy to recruit a large number of
participants, but hard to recruit high-quality participants.

Despite the warning signs in Y1, we continued the same free service model in
Y2---and experienced the same set of problems.
Luckily by Y3 we were already moving to a different payment model.
We would like to claim that this was done consciously to improve the quality
of our participant pool.
But in reality it happened because of a miscommunication with Sprint.
We had understood that they could provide individual liable plans to
participants at the discounted group rate that we had negotiated.
This turned out not to be the case.
Instead, we had to keep participants on the corporate liable plan and begin
handling billing ourselves.

Once we began charging participants a few things changed immediately.
First, we had to work harder to recruit participants.
That was expected.
But second, we began to see a large increase in participants porting their
existing number into the project.
This turned out to be a very good sign, and something that we should have
looked for earlier.
It indicated that participants were moving their existing cellular identity
on to their \PhoneLab{} smartphone, a clear signal that it was now their
primary device.

\begin{figure}[t]
	\centering
	\includegraphics[width=0.48\textwidth]{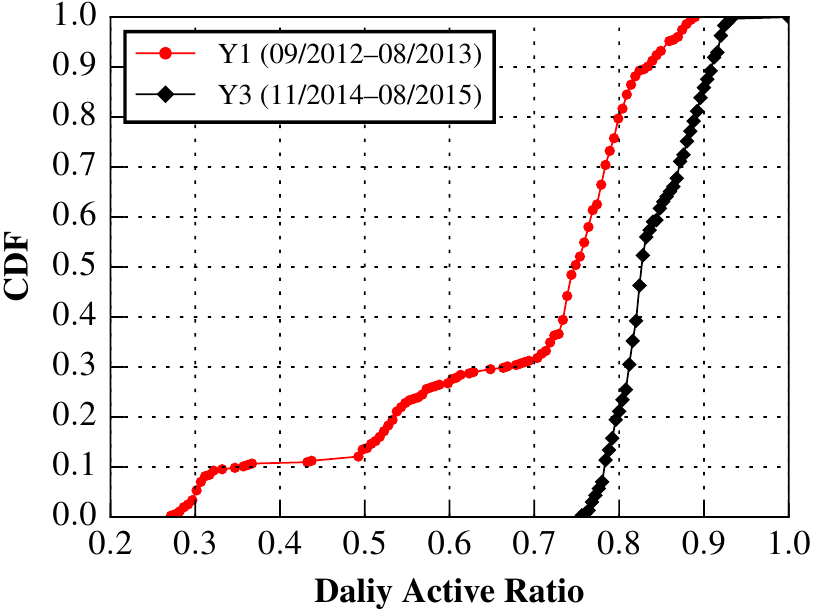}
	\caption{\textbf{CDF of Daily Active Ratio for Y1 and Y3.}
    Days when the backend server was down are not counted.
	Switching from free (Y1) to paid service (Y3) improves the ratio of active
	devices.}
	\label{fig:ratio}
  \vspace*{\aftercaptionskip}
\end{figure}

To examine the effectiveness of the incentive adjustment, we compare the
ratio of daily active devices before and after the change.
The ratio indicates the percentage of participants that are using their
device daily, and probably as their primary device.
Figure~\ref{fig:ratio} shows the CDF of daily active ratios for Y1 and Y3.
We can see that participants who paid for service (Y3) are more likely to use
the device than people who received service for free (Y1).
The daily active ratio for Y3 consistently falls in the 80\%--90\% range.

Getting the incentives right was probably our biggest challenge and getting
them wrong our biggest mistake.
While it all seems obvious in hindsight, we burned through two years and a
large amount of funding before assembling a high-quality participant pool.
In retrospect, there were several compounding factors that contributed to
this error.
First, like most computer scientists, we were overly focused on system
building and neglected the human aspects of the project.
This myopia may have also effected proposal reviewers and others that
commented on the project at early stages.
Despite clear problems with our incentive design, we cannot recall anyone
raising concerns about that part of the project.

Second, our proposal was set up to pay for service fees.
We went to a lot of trouble to get those allocations approved, and it would
have been difficult to reuse them for something else.
In today's funding climate, nobody ever wants to turn away funding.
But we probably would have needed to in order to reconfigure the award to
remove the service fees.

Finally, we had set an aggressive growth target of reaching 1000~\PhoneLab{}
participants in three years.
Today that number seems both ludicrous and unnecessary.
There is great value in being able to move an experiment out of a small-scale
lab setting and on to 100 participants.
But at that point, the marginal improvement in moving to 1000 participants is
limited.
We have never had an idea that we could not evaluate because \PhoneLab{} did
not have 1000~participants.
Nor has any experimenter ever expressed a need for that scale.
But in the early years, faced with a need to grow the project and the funds
available to do it, it was easy to stick with the option that made it easy to
recruit participants---even if they were low-quality participants.

\subsubsection{Emphasize Non-Monetary Incentives}
\lesson{Monetary incentives are expensive.}

The monthly price that we currently charge participants is not much lower
than what they could get on the open market.
In fact, most family plans are usually cheaper, and so most of our
participants are not in a position to join a family plan.
As a result, to incentivize participants to join we have had to construct and
emphasize non-monetary incentives.

One that has been simple and effective is highlighting the fact that
\PhoneLab{} offers friendly and convenient on-campus technical support.
We advertise drop-in office hours where participants can come by and receive
help with their \PhoneLab{} smartphone.
Common reasons for visits include broken or lost devices, configuration
problems, or performance concerns.
This turns an irritation---we have to provide first-level customer
service---into a selling point.

Given the compact layout of our university's campus, our lab is easier for
participants to visit than nearby Sprint stores.
And the service they receive is both different and
better than what they would receive in a store.
Our current \PhoneLab{} administrator is more technically savvy and
resourceful than your average cellular store employee.
Unconstrained by corporate policies, he is free to focus on providing the
best service possible to participants.
\begin{comment}
Our administrator has a background in the community dedicated to modding
smartphones to accomplish what they want the device to do. This has given
him a very unique knowledge and skillset when working with these devices.
Oftentimes a participant has come in with a device that is either
completely dead or in an unusable state and he has been able to revive
the device to the degree that a participant is able to continue using
their device, or be able to retrieve the data from the device and move
that data over to a new device. This was invaluable when we were unable
to purchase more nexus 5 devices for replacements. Our administrator began
to take apart devices that had smashed screens, but completely usable parts,
in order to increase the longevity of the devices that participants were using.
He also had a variety of techniques available to him that would allow him to
pull data off of a device provided he could get the device to turn on. In some
cases this involved taking the motherboard out of one device and placing it in
another device. These unorthodox techniques allowed him to provide support to
the participants in ways that are not possible for an ordinary cellular store
employee. Having this flexibility for the administrator was what made this level
of service possible.
\end{comment}

\subsubsection{Don't Be Afraid to Drop Participants}
\lesson{When 5\% of participants cause 95\% of your problems, you can easily
eliminate 95\% of your problems.}

Paying participants have greatly improved the quality of the \PhoneLab{}
testbed.
But when you start charging participants, expectations change.
Not only do we now have to handle billing and manage cash flow, but we also
act as a first point of service for \PhoneLab{} participants.

Not all participants are created equal.
Some pay their bills on time, read your emails, come in for help at
designated times only when they have a serious problem, and interact well
with your staff.
Others miss every billing cycle, ignore your emails, drop by your lab at any
time with any kind of minor smartphone irritation, and verbally abuse your
staff. One participant in particular made it a point to call our administrator
at late hours to tell him his phone is having issues and that the administrator
was putting his patients at risk as a result of the device not working
properly.

After removing the free service incentive, we have had to work hard to
recruit participants willing to pay to be a part of a research project.
But over time, we have realized that it can be helpful to encourage some to
leave the project.
Given that we are trying to optimize limited staff resources, retaining
participants with above average service expectations is not worth it.
Over the past two years, our \PhoneLab{} administrator has been given
permission to quietly suggest to some participants that they might be happier
with another service provider.
And removing the heavy hitters makes it easier to provide acceptable service
to remaining participants.

\begin{comment}
\subsubsection{Most Users Are Great}
\lesson{The other 95\% of participants are patient and understanding.}

Once you remove the participants that generate large numbers of issues and
complaints, the rest are fine and even sometimes fun.
We have had many positive interactions with participants throughout the
course of the project.
One lady has limped across campus on her newly surgically-repaired knee but
arrived just after our administrator had left for lunch.
Although she considered beating him with her cane, she instead decided to
come back later.
Another gentleman was kind enough to purchase pizza for the entire lab after
a week in which he needed and received service several times.
\end{comment}
 \subsection{Experimentation}

Our second set of lessons concerns how to build a testbed that best supports
useful experimentation.
Just as \PhoneLab{} is nothing without participants, it is useless without
experimenters.
The goal here is similar to the goal when building a participant base:
maximize the use of the testbed given a fixed set of resources.
With this goal in mind, we offer the following recommendations.

\subsubsection{Eat All of Your Own Dog Food}
\lesson{Testbed developers should be testbed experimenters and, if possible,
testbed participants.}

From the beginning, \PhoneLab{} developers have been both experimenters and
participants.
As experimenters, they have a direct interest in ensuring that the testbed is
usable and effective.
\PhoneLab{} developers have contributed much of the current platform
instrumentation in support of their own research projects.
They have also contributed new experimental features, such as integration
with the Linux kernel logging subsystem, again in support of their own work.
\PhoneLab{} developers that are also experimenters have stronger
incentives to contribute to the project than a developer paid to implement
functionality needed by others.

Simultaneously, \PhoneLab{} developers are also participants.
The developers form a subset of participants that receive experimental
changes first.
This is an ideal arrangement.
\PhoneLab{} developers are the first group to feel the impact of the changes
resulting from a new experiment.
And because they are more tech savvy than other participants, they are more
sensitive to problems that new experiments may be causing.
Developer testing has stopped several experiments from continuing to the full
testbed, including extremely obvious regressions (every app starts crashing)
and unacceptable performance degradations.

\subsubsection{Fly Under the Radar When Needed}
\lesson{Asking for forgiveness is better than asking for permission. But
don't ask, don't tell is best.}

\PhoneLab{} would not be possible without the excellent working relationship
we have with Sprint.
Cellular providers have a reputation of being difficult to work with.
Consider how many research papers describe the performance of ``Carrier A''
or ``a major US cellular carrier'', apparently to avoid jeopardizing a
working relationship with an oversensitive company.
So many researchers have inquired about how we are able to work so
effectively with Sprint.

Luck certainly played a role.
We were fortunate to have initial contacts at Sprint that may have been
unusually open-minded for the telecom industry.
But, more importantly, our initial contacts with Sprint were in sales, not
technology or network infrastructure.
Our experience has been that the sales side of Sprint is concerned with how
many lines we are purchasing and unconcerned with what we are doing with
them.
We have never concealed what we are doing from Sprint.
We have made it clear to them that we are deploying experimental Android
platform images to unlocked devices that connect to their network.
But we have had the good fortune to be telling the right people.
Had we told someone with a deeper understanding of exactly what that meant,
it is possible that \PhoneLab{} never would have existed.

\subsubsection{Go Where You're Needed}
\label{subsec:purpose}
\lesson{A good testbed does something that you can't do any other way.}

When \PhoneLab{} was designed and initially funded it was intended to support
both smartphone app and platform experimentation.
Because it was easier and safer, work began on app experimentation first, and
\PhoneLab{} initially opened with only this capability available.

But it quickly became apparent to us that \PhoneLab{} was neither the only
nor the best way to perform app experiments.
Deploying an app on existing software marketplaces and encouraging people to
install it is a much better approach.
Part of this is due to the different distribution model available for apps.
But it is also due to the realities of app experiments.
\PhoneLab{} could be used to force apps to be installed on participants'
smartphones---but we cannot force anyone to \textit{use} the app, or to use
it appropriately.
That requires a separate set of incentives, or an app that provides obvious
benefit to users.
Once you solve that problem, the benefit provided by \PhoneLab{} is minimal.

At that point we realized that the platform instrumentation and
experimentation capabilities were going to become much more important.
We quickly shut down app experimentation entirely and focused all of our
efforts on be able to do something that researchers could not do in any other
way.
Keeping a testbed relevant requires continuously surveying the experimental
landscape and identifying a useful niche.
To this day, \PhoneLab{} remains the only way to deploy platform changes and
collect data from several hundred representative smartphone users.

\subsubsection{Don't Be Too Afraid}
\lesson{Fear is the enemy of user-facing experiments.}

We have always been aware that an errant \PhoneLab{} experiment could cause
serious problems for our participants.
The ultimate nightmare scenario is a platform update that is so broken that
participants have to actually return to the lab for a fix.
This would not only represent a nuisance, but a real safety problem for
participants that rely on their smartphones in an emergency.

Happily, that has never happened.
But we have pushed updates with severe problems.
An initial attempt by external researchers to instrument the SQLite database
used by Android apps generate an unhandled exception.
That caused all Android apps that use SQLite---the vast majority---to crash
fairly quickly after launch.
Of course, this problem was noticed immediately by our developer testers and
a fix followed the initial regression within 24~hours.
Luckily, the \PhoneLab{} Conductor that is required to update the platform
does not use SQLite and so was not affected by this bug.
Had that not been the case, the problem would have been much more severe.

It is important to recognize the potential consequences of poorly-designed
experiments when running a human-facing testbed.
But it is equally important that that concern not hold back interesting
experiments.
As we gained experience with platform updates, we overcame initial concerns
about systemic failures.
We also gained confidence in our ability to catch severe regressions through
pre-deployment testing.

\subsubsection{Offload Human Subjects Review}
\label{subsec:irb}
\lesson{The IRB is an easy human solution to potentially hard or
time-consuming technical problems.}

Privacy concerns have always been a part of operating \PhoneLab{}.
This is particularly true once we began distributing platform experiments.
Android has a permission mechanism that limits app's access to sensitive
information.
This is useful when distributing app experiments.
But platform code is entirely trusted and runs unconstrained by Android's
permission mechanisms.
Distributing platform changes opens the door to all kinds of malicious
experiments, which have full access to any and all user data, including
things that apps cannot even request access to.
For example, it is entirely possible for a platform change to collect and
offload a full-resolution video feed harvested from the
\texttt{SurfaceFlinger} screen buffer.
Or implement a key logger to collect user passwords and other sensitive
inputs.

We considered multiple ways to try to ensure that \PhoneLab{} experiments
were safe for participants.
We could have tried to analyze experiments using static or dynamic analysis
to determine whether they matched the description provided by the
researchers.
It is not clear whether this is technically feasible, and would have been
unfamiliar terrain for our group which does not study computer security.
We also could have inspected them ourselves be examining the source code
changes required by the experiment.
This would have been time consuming and error prone.

Instead, we fell back on a simpler approach---require that researchers clear
their experiment with their Institutional Review Board (IRB) or equivalent.
Not all experimenters appreciated this mandate, and we had several cases
where interest evaporated once we explained the IRB requirement.
And we are definitely aware of frustration and even disdain for human
subjects review within the mobile systems community.
We have heard the IRB described as a nuisance, and even as something that
researchers do as busy work to avoid doing actual research.

But from our perspective the IRB is an ideal solution.
Note that we do not believe that most IRBs will adequately review the
experiment itself---we could probably do a better job by examining the diffs.
The point is not to improve how the experiment is reviewed, the point is to
establish a chain of responsibility that avoid us but leads back to the
experimenter's creators.
If we review an unsafe experiment incorrectly, then we are partly at fault.
If the review process is done by others, we can avoid worrying about
culpability.

Overall we believe that the mobile systems research community needs to
embrace the IRB process to help protect the work we do with human subjects.
It is true that the IRB at many institutions is disturbingly slow and poor at
reviewing computer science experiments.
But that is not a reason to avoid it---it is an argument for fixing it.
And in many cases, \PhoneLab{} experiments have qualified for expedited IRB
review due to their limited impact on human subjects.

It is also important to understand when IRB approval is required and when it
is not.
Preliminary analysis of data sets that does not reveal information in the
form of a publication does not require IRB approval.
This allows us to provide experimenters with previews of data sets generated
by their experiments or other \PhoneLab{} instrumentation.
If the data is useful and they want to proceed to publication,
they can initiate the IRB approval process at that point.
 \subsection{Testbed Development and Operation}

\PhoneLab{} and other similar testbed are usually run by researchers but are
not research projects.
As a result, they produce a unique set of development and operational
challenges that researchers are not always prepared for.
Unlike a research prototype, which only has to work until the paper deadline,
testbed software has to work reliably for long periods of time.
This is particularly true for testbeds like \PhoneLab{} that have human
participants who are generally intolerant of problems and failures.
Testbed development is also a great deal of work when measured
against certain research incentives---such as publications about the testbed,
rather than those that are facilitated by it.
And operating human-facing testbeds requires continuous effort and attention.
It is not acceptable for the entire testbed to fail for even one day, given that
human participants may be affected.
Given these requirements, we offer the following recommendations.

\subsubsection{Do the Minimum}
\lesson{Don't build anything you don't desperately need.}

\PhoneLab{} was initially extremely overdesigned.
There were multiple reasons for this.
It is natural to design large and complex ``proposal-ware'' to try to attract
funding.
It is also tempting to create new software components to ease the management of
a large pool of participants.
But regardless of the reasons, we began the project with plans to build many
different pieces of software that we ended up not needing.
Some of them we wasted time building anyway---others were fortunate enough to
slip off of the end of the queue.
We estimate that the parts of the \PhoneLab{} software base that are heavily
used---over-the-air updates and log data collection---probably represent
around 25\% of the overall development effort.
Not surprisingly, these are the components that are the most immediately
obvious and that we would have desperately needed on day 1.

As an example of an unnecessary feature, we initially implemented a heartbeat
mechanism.
Periodically \PhoneLab{} devices push small pieces of structured data back to
our central server.
This data is consumed and used to update a variety of different database
tables storing metadata about each connected device.
However, this information was never frequently used nor completely
ignored.
It had some limited utility when we were operating our free service plans,
since we tried to use it to determine users' activity levels.
But since we have moved to paying participants we have not needed and no
longer used this information.
Another example is the configuration system for the \PhoneLab{} Conductor.
We built a fairly complicated system for retrieving configuration parameters
from our servers and using them to reconfigure various Conductor components.
Most of these parameters, however, have never been changed, and overall the
system could have been designed in a much simpler way.

Almost all systems are overdesigned to some degree.
But given that \PhoneLab{} was implemented by faculty and students who were
also trying to do actual research, the wasted effort is more problematic.
Given the chance to start again, we would let our own needs as \PhoneLab{}
experimenters drive what needed to be built.

\subsubsection{Small, Developer-Heavy Teams Are Best}
\lesson{Testbed development requires many lines of code and few if any new
ideas.}

A typical systems research project that faculty are used to leading involves
the exploration of many new ideas through software prototypes.
Testbed development is different.
It involves the reliable implementation of a few simple ideas---less
brainstorming and more testing---less creativity and more coding.
As a result, it requires a different kind of team than would be assembled to
conduct a research project.

Our experience has been that too many faculty trying to lead a testbed
development project tend to slow things down.
They tend to come up with ideas faster than available developers can
implement them, and have a hard time focusing on the core tasks that need to
be done well.
In a research project having a bunch of students chasing after a bunch of
ideas is fine and even productive.
In a testbed project, without new ideas to explore, it's more important to
focus on doing a few things well.
As funding for \PhoneLab{} dried up, our team shrunk naturally---from five
active faculty and three full-time graduate students; to a single faculty
member, one part-time graduate developer, and one part-time undergraduate
administrator.
Far from suffering as staffing was reduced, the testbed actually benefited
from a lower decision maker to developer ratio.
A lean project team also helps avoid building unnecessary things, as
described previously.
  \section{Whither Smartphone Testbeds?}
\label{sec:discussion}

Public testbeds have a rich history in the systems and networking community.
EmuLab, PlanetLab, and MoteLab helped accelerate research in networking,
planetary-scale systems, and wireless sensor networks.
\PhoneLab{} has successfully supported multiple research experiments and
generated a great deal of useful data.
We hope that it has had a beneficial effect on the mobile systems community.

But our experience with \PhoneLab{} has made us concerned about the future of
this kind of shared infrastructure.
Public testbeds require three types of support to succeed.
First, they need initial and continued funding.
Second, other incentives have to exist encouraging people to build and
operate them.
Third, they need community buy-in through active experimentation and norms
that encourage and require experimentation at scale to validate new ideas.
Today, all three of these sources of support seem increasing uncertain.
We discuss each in turn below.

\subsection{Testbed Funding}

At least in the United States, the days where it was easy to get and keep
testbeds funded seem behind us.
For example, the EmuLab networking testbed received regular infusions of
funding from the NSF.
Neither PlanetLab nor MoteLab were as successful as EmuLab at obtaining
funding, and neither were we.
A move toward smaller and more focused awards may make it more difficult to
make the significant investments required to build community infrastructure.

User-facing testbeds face their own unique funding challenges.
Equipment-based testbeds have large up-front costs and low continuing costs.
The testbed funding programs that we are familiar with are set up to reflect
this, and often place limits on the amount of operational support that can be
included.
In contrast, user-facing testbeds can have both large up-front costs and high
continuing costs.
Continuing to operate \PhoneLab{} requires not only interacting with
experimenters, but continuing to handle billing and service requests for
\PhoneLab{} participants.
Grants for testbeds have also grown shorter, which also challenges
user-facing projects.
It took us several years to build up a suitable participant base, by which
point our funding had almost expired.

We have considered other funding models.
We are not in the position to establish a consortium of the kind that
successfully funded PlanetLab.
Nor do we think that it would be a successful model for smartphone platform
experimentation.
The small number of companies with the ability to deploy Android platform
modifications have the ability to experiment on millions of users.
So this is quite a different marketplace than the somewhat larger number of
companies with an interest in building planetary-scale systems.

We have also considered charging groups to use \PhoneLab{}.
But this runs counter to an established tradition in the research community
of offering shared infrastructure free of charge.
We also did not want to create additional barriers for groups willing to
perform the difficult task of modifying Android.

We hope that funding agencies will consider the challenges of user-facing
systems as they plan the next generation of mobile system testbeds.
There are exciting efforts underway in the US to build city-scale testbeds
that combine wide-area wireless networks with mobile and embedded devices.
These testbeds must integrate with smartphones and other user-facing devices.
Yes---that does make them much harder and more expensive to operate.
But cities of the future will still have people living in them---at least we
hope.

\subsection{Operator Incentives}

More significant public funding for earlier testbeds also helped create an
incentive for groups to build and operate them.
EmuLab grants, for example, were large enough not only to support testbed
operations, but also help support a successful networking research group.
But the move to smaller, shorter, and more focused grants removes this
incentive for groups to operate testbeds.

Financial incentives for operators are important, but so is community
recognition.
It is also much cheaper.
We hope that the mobile systems community will continue to publish papers
describing important tools and infrastructure, even if they do not contain
research discoveries.
From an efficiency standpoint, building a testbed is probably the worst way
to generate a research paper.
But if there is no way to publish papers about tools and testbeds, it further
drains motivation from their creators and maintainers.

\subsection{Community Buy-In}

Finally, research communities need to support testbeds by using them.
This is not as easy as ``If you build it, they will come.''
We have built \PhoneLab{}, made considerable efforts to publicize the
project, and make experimentation as simple as possible.
But we have never been satisfied the amount that \PhoneLab{} was used by the
mobile system research community.

Why didn't people use \PhoneLab{}?
It is true that modifying Android is hard, and no amount of instructions on
our part can mitigate the difficult of experiment-specific changes.
But experimenters have to make these changes anyway, regardless of whether
they deploy them on \PhoneLab{} or not.
We also do require that experimenters receive IRB approval, for reasons
explained earlier, and know for a fact that that has turned some away.
But at many institutions this is more of a nuisance than a good reason to not
do a large-scale experiment.
Given that AOSP is more of a code dump than an open source project, there is
also not an easy route for successful changes to make their way into actual
Android releases.
But this does not distinguish research done on \PhoneLab{} from many other
kinds of system building where it is the ideas, rather than the artifacts,
that may someday have an impact on a product.

We believe that the most important reason is that community has not demanded
that researchers perform larger-scale experiments.
To a certain degree, as long as researchers can get away with publishing
papers based on small-scale studies, they will continue to do so.
Running experiments on testbeds like \PhoneLab{} not only takes time and
energy, but also may challenge conclusions drawn from small-scale studies.
We regularly read papers that present results from small-scale studies that
could have been run on \PhoneLab{}.
Stronger community norms are needed to encourage researchers to make use of
available testbed facilities when appropriate.
 \section{Related Work}
\label{sec:related}

The \netsense{}~\cite{striegel2013lessons} project has been using smartphones
to study the social interaction between college students.
Free Nexus~S smartphones were distrusted to 200 university freshmen for two
years.
The smartphones ran modified CyanogenMod and instrumentation was added to log
the communication events---such as phone calls, SMS, Facebook posts and
Bluetooth proximity.
Unlike \PhoneLab{}, NetSense focuses on behavioral rather than systems
experiments.
The testbed is not open, nor is there a way to distribute platform
modifications.
At this point NetSense has moved to running as an app and utilizing a ``bring
your own device'' model, rendering it complicated or impossible to perform
platform experiments.

LiveLabs~\cite{misra2013livelabs,balan2014livelabs,jayarajah2016livelabs} is
a human behavioral experiment testbed utilizing smartphones.
They do not hand out smartphones nor control the platform, but instead deploy
experiment software on participants' own devices.
Because of this, LiveLabs is able to scale up to several thousands of
participants spanning three venues, including university campus, a resort
island and a large convention center.
LiveLabs has different aims than \PhoneLab{}.
Its goal is to enable more pervasive computing experiments, rather than work
on smartphone systems.

Finally, SmartLab~\cite{larkou2013smartlab} is a smartphone testbed
consisting of 40 Android smartphones.
The smartphones are connected to a hub via USB and user interactions are
simulated through a web-based remote screen terminal.
The devices are neither mobile nor used by real users.
\PhoneLab{} provides a level of realism that SmartLab lacks.

There are also various attempts to deploy experiments as apps on software marketplaces.
MobiPerf~\cite{huang2011mobiperf} is an Android app that utilizes the
Mobilyzer~\cite{nikravesh2015mobilyzer} library to perform network
measurements, such as bandwidth and latency testing.
The app was deployed on the Google Play store and has over 10K installations
so far.
Device Analyzer~\cite{wagner2013device} is a Android data collection tool
that collects various information at background, such as phone charging
status, phone calls, Bluetooth proximity, and so on.
Different with MobiPerf, Device Analyzer does not provide value as the app
and relies on voluntary participation.
Compared to app-based measurement tools, \PhoneLab{} has access to unfiltered
more detailed information by instrumenting the smartphone platform.

\section{Conclusion}
\label{sec:conclusion}

To conclude, we have described the \PhoneLab{} smartphone platform testbed.
We have provided a description of the testbed and an overview of how the
testbed works.
We have described example \PhoneLab{} experiments and provided an overview of
available \PhoneLab{} datasets.
We have attempted to distill some of the lessons that we have learned while
building and operating \PhoneLab{}, and discussed some of the implications of
our experiences for the next generation of mobile systems testbeds.

Unfortunately, funding for \PhoneLab{} will run out at the end of next year,
and so experimentation will cease in mid-February 2017.
We will continue to provide access to our data sets after that point, and
offer up our software for anyone with similar data collection needs.
Although \PhoneLab{} is ending in its current form, we are committed to
maintaining experimental access to this important codebase.
We plan to approach the CyanogenMod community to see if they are willing to
participate in a future incarnation of \PhoneLab{}.
Experiments that only include modders will lack the realism that our current
participant base provides.
But this approach may allow us to continue platform experiments at an even
lower overhead.

\clearpage
{
\footnotesize
\bibliographystyle{acm}

}
\end{document}